\documentclass[preprint,showkeys,tightenlines,amsmath,amssymb,12pt]{article}

\usepackage{graphicx}
\usepackage{amssymb}
\newcommand{\EkB}{\,\mathcal{E}/k_{\rm B}}
\newcommand{\calE}{\,\mathcal{E}}

\begin{document}

\title{Structural Basis of Folding Cooperativity in Model Proteins: Insights
  from a Microcanonical Perspective}

\author{Tristan Bereau and Markus Deserno\\
  Department of Physics, Carnegie Mellon University,\\
  Pittsburgh, PA 15213\\ \\
  Michael Bachmann\\
  Center for Simulational Physics,\\ Department of Physics and Astronomy,\\
  The University of Georgia, Athens, GA 30602
}

\date{\today}

\maketitle

\vspace{25mm}

Running title: Protein Folding Cooperativity\\ 

keywords: Protein folding, Coarse-grained simulations, Microcanonical
  analysis, Two-state cooperativity, Downhill folding, Helical
  peptides.\\ \\

\newpage

\begin{abstract}
  Two-state cooperativity is an important characteristic in protein
  folding.  It is defined by a depletion of states lying energetically
  between folded and unfolded conformations.  While there are
  different ways to test for two-state cooperativity, most of them
  probe indirect proxies of this depletion. Yet, generalized-ensemble
  computer simulations allow to unambiguously identify this transition
  by a microcanonical analysis on the basis of the density of states.
  Here we perform a detailed characterization of several helical
  peptides using coarse-grained simulations. The level of resolution
  of the coarse-grained model allows to study realistic structures
  ranging from small $\alpha$-helices to a \emph{de novo} three-helix
  bundle---without biasing the force field toward the native state of
  the protein.  Linking thermodynamic and structural features shows
  that while short $\alpha$-helices exhibit two-state cooperativity,
  the type of transition changes for longer chain lengths because the
  chain forms multiple helix nucleation sites, stabilizing a
  significant population of intermediate states. The helix bundle
  exhibits the signs of two-state cooperativity owing to favorable
  helix-helix interactions, as predicted from theoretical models. The
  detailed analysis of secondary and tertiary structure formation fits
  well into the framework of several folding mechanisms and confirms
  features observed so far only in lattice models.

\end{abstract}

\section{Introduction}

Two-state protein folding is characterized by a single free-energy barrier
between folded and unfolded conformations at the transition temperature
$T_{\rm c}$, whereas downhill folders do not exhibit folding barriers
\cite{bryngelson, dobson}. The analysis of this property conveys important
information on both the thermodynamics as well as the kinetic pathways of
proteins \cite{bryngelson, jackson98}.  A widely used test for a two-state
transition is the calorimetric criterion, which probes features in the
canonical specific heat curve \cite{Chan_kappa2}. However, this criterion
neither provides a \emph{sufficient} condition to identify two-state
transitions \cite{zhhaka99}, nor does it offer a clear distinction between
weakly two-state and downhill folders. Other experimentally observable aspects
of two-state cooperativity include sharp transitions in certain order
parameters, or features in chevron plots \cite{ChBrDi95, jackson98}. All these
methods focus on thermodynamic consequences of a depletion of intermediate
states; they don't study it directly.

However, it is possible to determine the density of states in a standard
canonical computer simulation at temperature $T^*$ of interest: sample the
probability density $p(E)$ of finding an energy $E$.  The density of states
$\Omega(E)$ is then proportional to $p(E)\,{\rm e}^{E/k_{\rm B}T^*}$, and
hence the entropy is (up to a constant) given by $S(E)=k_{\rm
  B}\ln\Omega(E)={\rm const.} + k_{\rm B}\ln p(E)+E/T^*$.  One may proceed to
analyze the system \emph{microcanonically}, i.e., to study the thermodynamics
of $S(E)$, in the neighborhood of $\langle E\rangle_{T^*}$.  The advantage is
that we essentially directly analyze the \emph{probability density} $p(E)$
rather than merely looking at its lowest \emph{moments}, such as the specific
heat.  Such a microcanonical analysis has been applied to a wide variety of
problems, e.g., spin systems \cite{Borgs1992, gross_book, huller, deserno97,
  Janke1998, HuPl02, PlBe05}, nuclear fragmentation \cite{Gross1993,
  Koonin1987}, colloids \cite{fernandez10}, gravitating systems
\cite{Komatsu09, Posch06}, off-lattice homo- and heteropolymer models
\cite{TaWoBi09, ChLiLiLi07}, and protein folding \cite{hao_scheraga, sikorski,
  JuBaJa06, junghans_JCP, HeGo08, keyes, bereau_jacs}.  Two remarks are
worthwhile:
\begin{itemize}
\item If the transition is characterized by a substantial barrier, standard
  canonical sampling suffers from the usual getting-stuck-problem: During a
  simulation the system might not sufficiently many times cross the barrier to
  equilibrate the two coexisting ensembles.  This, of course, is true and
  needs to be avoided irrespective of whether one aims at a canonical or
  microcanonical analysis.  Many ways around this problem have been proposed,
  e.g., multicanonical \cite{multicanonical} or Wang-Landau \cite{wanglandau}
  sampling.  In our study we employ replica-exchange molecular dynamics for
  sampling coupled canonical ensembles \cite{p_tempering} and combine the
  overlapping energy histograms by means of the weighted histogram analysis
  method (WHAM) \cite{FeSw89, kumar92, bereau_WHAM}, a minimum variance
  estimator for $\Omega(E)$.
\item Accurately sampling the whole distribution $p(E)$ over some
  range of interest requires better statistics than merely sampling
  its lowest moments: there's a price for higher quality data.  But
  then, a microcanonical analysis taps into this quality, while a
  canonical analysis of the much longer simulation run would not
  significantly improve the observables.  Recall that the canonical
  partition function $Z(T)=\int {\rm d}E\;\Omega(E) \exp^{-E/k_{\rm
      B}T}$ is the \emph{Laplace transform} of the density of states
  $\Omega(E)$, an operation well-known to be ($i$) strongly smoothing
  and thus ($ii$) hard to invert.
\end{itemize}
From a thermodynamic point of view a two-state transition is
characterized by two coexisting ensembles of conformations
\cite{ChBrDi95}. While this does not qualify as a genuine (first
order) phase transition (because the free energy of finite systems is
always analytical), its finite-size equivalent can be unambiguously
characterized by monitoring the entropy $S(E)$.  In the
phase-coexistence region it will exhibit a \emph{convex intruder} due
to the suppression of states of intermediate energy.  This can best be
observed by defining the quantity $\Delta S(E)=\mathcal{H}(E)-S(E)$,
where $\mathcal{H}(E)$ corresponds to the (double-)tangent to $S(E)$
in the transition region \cite{gross_book, huller, JuBaJa06,
  junghans_JCP, deserno97}. In a finite system the existence of a
barrier in $\Delta S(E)$ will imply a non-zero microcanonical latent
heat $\Delta Q$, defined by the interval over which $S(E)$ departs
from its convex hull, and in turn leads to a ``backbending'' effect
(akin to a van-der-Waals loop) in the inverse microcanonical
temperature $T_{\mu {\rm c}}^{-1}(E)=\partial S/\partial E$ (e.g.,
\cite{gross_book, huller, deserno97, JuBaJa06, HuPl02, PlBe05}). A
non-zero $\Delta Q$ demarcates a transition \emph{region}, whereas a
downhill folder (continuous transition) will only exhibit a transition
\emph{point}, where the concavity of $S(E)$ is minimal.

Extending a recent study \cite{bereau_jacs}, we focus here on the link
between ($i$) the nature of the transition (i.e., two-state vs.\
downhill), ($ii$) secondary structure, and ($iii$) tertiary structure
formation for several helical peptides using a high-resolution,
implicit-solvent coarse-grained model. The results will be interpreted
in terms of different frameworks of folding mechanisms, such as the
molten globule model and simple polymer collapse models
\cite{dill_stigter, Baldwin}.  While all helical peptides presented in
this work are artificially constructed (``{\sl de novo}''), and have
thus not naturally evolved, they exhibit the relevant physics in a
particularly clean way and are in this sense useful model systems.
(See Supporting material for a further discussion of this point.)

\section{Methods}

Coarse-grained (CG) Molecular Dynamics (MD) simulations were based on
an intermediate resolution, implicit-solvent peptide model
\cite{bereau_jcp}.  It accounts for amino acid specificity and is
capable of representing genuine secondary structure \emph{without}
explicitly biasing the force field toward any particular conformation
(native or not).  Table~\ref{tab:aa_seq} lists the sequences of all
studied peptides.  More details can be found in the Supporting
Material.

Replica-exchange MD simulations were performed using the {\sc
  ESPResSo} package \cite{espresso}. All simulations were run in the
canonical ($NVT$) ensemble using a Langevin thermostat with friction
constant $\Gamma=\tau^{-1}$, where $\tau$ is the intrinsic unit of
time of the CG model. The CG unit of energy, $\calE$, relates to
thermal energy at room temperature via $\calE = k_{\rm B}T_{\rm
  room}=1.38\times10^{-23}\,{\rm J}\,{\rm K}^{-1}\times300\,{\rm
  K}\approx 0.6\,{\rm kcal}\,{\rm mol}^{-1}$. The temperature $T$ was expressed in
terms of the intrinsic unit of energy $T=\EkB$. The equations of
motion were integrated with a time step $\delta t=0.01\,\tau$.

Entropy, order parameters, and canonical averages were obtained from
the density of states, $\Omega(E)$, which itself was calculated from
WHAM \cite{FeSw89, kumar92, bereau_WHAM}.  Details can again be found
in the Supporting Material.

Finally, the reader should observe that CG force fields---including
the one used here---are usually constructed to reproduce the canonical
ensemble, hence they strive to reproduce the free energy.  However,
individual enthalpic and entropic contributions will generally be off,
because the reduced number of degrees of freedom lowers the entropy of
CG conformations, and so the energies need to be adjusted to leave the
free energy correct.  For instance, in the absence of solvent both
solvent energy and entropy must be parametrized into effective solute
interaction energies.  The entropies we calculate in this work are
thus not to be confused with the entropies of the actual system.  On
the other hand, this does of course not deprive them of being
exquisitely sensitive observables for the thermodynamics of the CG
model.

\section{Results}

\subsection{Secondary structure}

We first examine the structural and energetic properties of the
sequence (AAQAA)$_n$ with various chain lengths $n=3,\,7,\,10,\,15$.
The $n=3$ variant is known as a stable $\alpha$-helix folder and has
been studied both experimentally and computationally \cite{scholtz91,
  shalongo94, zagrovic05, ferrara00, mousseau}.  The $n=7$ peptide has
also been shown to fold into a helix \cite{zagrovic05}. We find that
all four peptides form a stable long helix in the lowest energy sector
(see below), but are not aware of any structural study that would
confirm this for the longer peptides with $n=10,\,15$.  Since we will
soon show that the latter two fold differently from the shorter ones,
an experimental confirmation of their ground state structure would be
very useful.

For (AAQAA)$_3$ Fig.~\ref{fig:entropy_n}a shows a barrier in $\Delta
S(E)$ as well as a backbending in the inverse microcanonical
temperature $T_{\mu {\rm c}}^{-1}(E)$, indicative of a first-order
like transition. The two vertical lines mark the transition region
with the corresponding microcanonical latent heat $\Delta Q$. In the
region between $E=(40-80)\,\calE$ mostly-helical and mostly-coil
conformations coexist, in agreement with the sharp transitions in the
helicity $\theta(E)$ (as determined by the {\sc stride} algorithm
\cite{stride}) and the number of helices in the chain, $H(E)$.  These
results point to a clear two-state folder.

Increasing the chain length from $n=3$ to $n=15$ (Figures
\ref{fig:entropy_n} b, c, d) changes the nature of the transition
significantly.  While $n=7$ still shows a (lower) barrier in $\Delta
S(E)$ and a non-zero microcanonical latent heat $\Delta Q$, the cases
$n=10$ and $n=15$ are downhill folders (no barrier in $\Delta S(E)$
and monotonic $T_{\mu {\rm c}}^{-1}(E)$ curves).  The transition
region is replaced by a transition point for which the concavity of
$S(E)$ is minimal and $\Delta Q=0$. This process is associated with
important structural changes around the transition region/point as
seen in the number of helices $H(E)$: while the curve is monotonic for
$n=3$, it exhibits a peak with $H(E) > 1$ for bigger $n$, showing that
during the transition most conformations form more than one helix.
This suggests the existence of multiple helix nucleation sites upon
folding (see representative conformations at the transition point for
$n=10,\,15$ in Fig.~\ref{fig:entropy_n}).

In order to further elucidate the structural features of these chains
around the transition region/point the fraction of secondary structure
(i.e., helicity) was analyzed in dependence of both energy and residue
index for helices $n=3,\,15$. While for $n=3$ helix nucleation appears
mostly around the center of the peptide and propagates symmetrically
to the termini (Fig.~\ref{fig:2d_hel_n}a), $n=15$ shows two distinct
peaks at an energy $E$ slightly below the transition point
(Fig.~\ref{fig:2d_hel_n}b). The results suggest the formation of two
individual helices placed symmetrically from the midpoint of the
chain---around residue 35---which only join into one long helix
significantly below the transition point. As will be discussed in
Section \ref{sec:discussion}, these two helices divide the system into
two distinct melting domains which fold non-cooperatively (i.e.,
folding one helix does not help folding the other) \cite{Privalov1989,
  Privalov1982}.  The same conclusion can be drawn from the
probability distributions of forming an $m$-helix (see Supporting
Material).

To probe the behavior of simultaneous folding motifs within a chain,
we performed a microcanonical analysis of the 73 residue \emph{de
  novo} three-helix bundle $\alpha$3D \cite{2a3d} (amino acid sequence
given in Table \ref{tab:aa_seq}). The CG model used here has been
shown to fold $\alpha$3D with the correct native structure, up to a
root-mean-square-deviation of $4\,{\rm \AA}$ from the NMR structure
\cite{bereau_jcp}. While of similar length compared to (AAQAA)$_{15}$,
it shows a \emph{discontinuous} transition (see
Fig.~\ref{fig:entropy_a3d}) and thus a nonzero microcanonical latent
heat during folding.  In the transition region the helicity increases
sharply from 20\% to about 65\%, and the average number of helices
also increases sharply -- but monotonically -- from 1.5 to 3. Unlike
for the simple $n=7,\,10,\,15$ helices, the transition region never
samples more helix nucleation sites than the number of helices at low
energies. As can be seen from the representative conformations shown
in Fig.~\ref{fig:entropy_a3d}, the ensemble of folded states
($E\approx 130\,\calE$) consists of three partially formed helices in
largely native chain topology; the coexisting unfolded ensemble
($E\approx 225\,\calE$) consists of a compact structure containing
transient helices.  All these findings identify $\alpha$3D as a
two-state folder.

To better monitor the formation of individual helices, we measured the
fraction of helicity as a function of energy and residue, see
Fig.~\ref{fig:2d_hel_a3d}. Unlike (AAQAA)$_n$ (Figure
\ref{fig:2d_hel_n}), $\alpha$3D shows strong features due to its more
interesting primary sequence. The turn regions (dark color) delimiting
the three helices (light color) are clearly visible at low energies
and correspond well to the {\sc stride} prediction of the NMR
structure, as shown in Table \ref{tab:aa_seq}. Moreover, it is clear
from this figure that secondary structure formation happens
simultaneously (i.e., at the same energy) for all three helices, and
that most of the folding happens within the coexistence region (marked
by the two vertical lines). The residues which form the native turn
regions do not show any statistically significant signal of helix
formation at any energy. Secondary structure has almost entirely
formed close to the folded ensemble in the transition region
(left-most vertical line)---in line with the representative
conformations shown in Fig.~\ref{fig:entropy_a3d}.

\subsection{Tertiary structure}

A secondary structure analysis alone can only provide information on
the local aspects of folding. Several studies have highlighted the
role of an interplay between local and non-local interactions in
protein folding cooperativity (see e.g.\ Refs.~\cite{kaya_chan_prl,
  ghosh_dill, bahato09, bereau_jacs}). Here we first analyze the size
and shape of the overall molecule by monitoring, respectively, the
radius of gyration $R_{\rm g}=\sqrt{\lambda_x^2 + \lambda_y^2 +
  \lambda_z^2}$ and the normalized acylindricity $c = (\lambda_x^2 +
\lambda_y^2) / 2\lambda_z^2$ as a function of $E$, expressed in terms
of the three eigenvalues of the gyration tensor $\lambda_x^2 <
\lambda_y^2 < \lambda_z^2$. The results for the single helices $n=3$
and $n=15$ and the three-helix bundle $\alpha$3D are shown in Figure
\ref{fig:gyr}. (AAQAA)$_3$ shows sharp features in both order
parameters within the transition region, indicating an overall
structural compaction (in shape and size) of the chain as energy is
lowered. Observe that $c$ approaches 0.13 at high energy, which is
close to the random walk or self-avoiding walk values, both close to
$c\approx0.15$ \cite{solc_rw, SAW_eig}. The longer helix $n=15$ shows
a non-monotonic behavior in both $R_{\rm g}(E)$ and $c(E)$: while the
radius of gyration exhibits a minimum around $E=400\,\calE$, the
normalized acylindricity displays a maximum. This indicates a
structure that is most compact and spherical $100\,\calE$ \emph{above}
the transition point. This dip in $R_{\rm g}(E)$ corresponds to a
chain collapse into ``maximally compact non-native states''
\cite{dill_stigter} due to a non-specific compaction of the chain
gradually restricted by steric clashes, at which point secondary
structure becomes favorable. Upon lowering the energy, the radius of
gyration increases and the acylindricity decreases, because the
peptide elongates while folding from a compact globule into an
$\alpha$-helix.  Results for the three-helix bundle are similar:
$R_{\rm g}(E)$ and $c(E)$ also show a minimum and a maximum,
respectively, slightly above the transition region. This indicates a
similar type of chain collapse mechanism. However, non-monotonic
features appear also at the other end of the transition region
($E\approx130\,\calE$) where the radius of gyration shows a
\emph{maximum} and the acylindricity plateaus. The evolution of the
two order parameters below the transition region is rather limited,
suggesting that only minor conformational changes take place (i.e.,
the shape of the molecule stays steady while its size decreases
slightly). In contrast, at high energy both (AAQAA)$_{15}$ and
$\alpha$3D are still far away from a random walk limit, as evidenced
by the acylindricity being far away from 0.15.

Chain collapse in longer chains (such as (AAQAA)$_{15}$ and
$\alpha$3D) can readily be observed by monitoring tertiary contacts as
a function of energy. Figure \ref{fig:1d_tert_a3d} shows the total
number of non-local contacts (red curve) as well as the number of
native contacts alone (blue curve). Tertiary contacts are defined here
as pairs of residues that are more than five amino acids apart (this
prevents chain connectivity artifacts) and within a $10\,{\rm \AA}$
distance (these numbers are somewhat arbitrary, but their value does
not affect the qualitative behavior of Fig.~\ref{fig:1d_tert_a3d}).
\emph{Native} contacts correspond here to the set of abovementioned
tertiary contacts sampled with a frequency higher than $1\%$ from a
set of $10,000$ low-energy conformations ($E\leq50\,\calE$). While the
two curves are virtually identical below the transition region (i.e.,
all contacts are native) and of similar trend above it, they behave
very differently \emph{inside} that interval. Although the number of
native contacts monotonically increases as the energy is lowered
(transition from globule to native-like structure), the total number
of contacts shows a peak above the transition region and sharply
decreases inside it.  To approach the native state, the peptide needs
to break more contacts of non-native type than it gains contacts which
are native.

The non-monotonicity of this curve, as well as the $R_{\rm g}$ data,
invite a comparison with the thermodynamics of water: upon cooling,
liquid water expands below 4$^{\circ}\,$C. Weak but isotropic van der
Waals interactions are given up for strong but directional hydrogen
bonds. This energy/entropy balance seems to occur in a very similar
manner here, and essentially for the same reason.  Weak van der Waals
side-chain interactions (i.e., tertiary contacts) are replaced by
hydrogen-bond interactions (i.e., secondary structure) at lower
energies. This further confirms the concept of a chain collapse into
maximally compact non-native states: upon lowering the energy (above
the transition region) the system has accumulated a large number of
non-native contacts due to a simple hydrophobicity-driven compaction
mechanism. This idea was proposed early on as the ``hydrophobic
collapse model'' or ``molten globule model'' \cite{dill_stigter,
  Baldwin}. A similar effect was observed by Hills and Brooks using a
G\=o model, where out-of-register contacts had to unfold in order to
reach the native state \cite{HillsJr2008}.

While a transient chain collapse upon cooling is present in both
(AAQAA)$_{15}$ and $\alpha$3D ($R_{\rm g}(E)$ is non-monotonic, see
Fig.~\ref{fig:gyr}b and c), its effect on tertiary structure formation
will greatly depend on the amino acid sequence.  Figure
\ref{fig:2d_tert} shows the number of tertiary contacts of the two
peptides as a function of energy and residue. The single helix $n=15$
shows a uniformly small number of tertiary contacts in the low energy
region (due to the linearity of the helix) and peaks {\em above} the
transition point (which corresponds to the energy where $R_{\rm g}(E)$
is smallest).  The tertiary contact distribution in the maximally
compact non-native states is homogeneous along the chain (i.e., all
residues have the same number of contacts). On the other hand, the
number of tertiary contacts along the three-helix bundle
(Fig.~\ref{fig:2d_tert}b) is highly structured, forming stripes as a
function of residue that extend below the transition region.  This
follows directly from the amphipathic nature of the subhelices that
constitute $\alpha$3D: residues that form the native hydrophobic core
of the bundle have a higher number of contacts. The presence of these
stripes in the energetic region of collapsed structures ($E\approx
300\,\calE$) is due to a strong selection between hydrophobic and
polar amino acids during the hydrophobic collapse, burying hydrophobic
groups inside the globule. The low number of tertiary contacts in the
turn regions indicates that they remain on the surface of the
maximally compact globule during chain collapse.

\section{Discussion}\label{sec:discussion}

Two-state cooperativity has been characterized as a common signature
of small proteins for which the transition of the cooperative domain
corresponds to the whole molecule (i.e., the protein undergoes a
transition as a whole) \cite{Privalov1979}. While this framework
applies well to the small helix (AAQAA)$_3$, it is difficult to
predict its thermodynamic signature from other grounds: a description
of the conventional helix-coil transition is not appropriate due to
the small size of the system and the correspondingly important
finite-size effects.

The thermodynamic signature of proteins can better be described for
longer chains. Several arguments can be brought forward to explain the
transition we observe for the longer helices (AAQAA)$_n$ for
$n=10,\,15$:
\begin{enumerate}
\item Most theoretical models of the helix-coil transition, such as
  the Zimm-Bragg model \cite{zimm1959}, are based on the
  one-dimensional Ising model, which -- being one-dimensional -- shows
  no genuine phase transition but only a finite peak in the specific
  heat.  The entropic gain of breaking a hydrogen-bond (i.e., forming
  two unaligned spins) outweighs the associated energetic cost for a
  sufficiently long chain.
\item The structure of the maximally compact state right above the
  transition (see Fig.~\ref{fig:2d_tert}) indicates that there is no
  statistically significant competition between amino acids (i.e., all
  residues have the same number of tertiary contacts) and is therefore
  associated with a homopolymer-type of collapse, which is indeed
  barrierless \cite{dill_stigter,Tiktopulo1994}.
\item The denaturation of large proteins composed of several
  ``melting'' domains is not a two-state transition
  \cite{Privalov1982, Privalov1989}. The presence of two helix
  nucleation sites around the transition point
  (Fig.~\ref{fig:2d_hel_n}) indicates the existence of two such
  melting domains that fold \emph{non}-cooperatively: folding one
  helix is not correlated with the formation of the other.  We have
  checked that there are no statistically significant helix-helix
  interactions between the two domains by calculating contact maps.
  These were averaged over the ensemble of conformations for which $50
  \le E \le 150\,\calE$ (data not shown).
\end{enumerate}
Common expectation is that bigger systems show sharper transition
signals, and it might thus appear surprising that the transition of
the (AAQAA)$_n$ sequence weakens for increasing $n$.  However, one
needs to bear two things in mind.  First, size alone is not
sufficient, dimensionality counts as well.  In the Supporting Material
we show examples of quasi-one-dimensional systems for which
transitions become weaker for bigger systems, because in the process
of growing they become ``more one-dimensional.''  When size is
associated with cooperativity, one tends to think of globular
(three-dimensional) systems, for which the size-cooperativity
connection is true, but this is not the most general case.  And
second, the sharpness might depend on what observable one studies.
The helicity $\theta$ as a function of temperature indeed varies more
sharply for larger $n$, making the response function
$(\partial\theta/\partial T)_n$ peak more strongly for bigger $n$.
While this steepening would suggest a stronger two-state nature, this
goes against every other observable which suggests a downhill
folder---including the calorimetric criterion (see below); observing
response functions alone can thus be misleading.  More details on this
can be found in the Supporting Material.

The two-state signature of the helix bundle $\alpha$3D can be
understood from two different perspectives:
\begin{enumerate}
\item While there are clearly three distinct folding motifs (i.e.,
  three helices), the selective hydrophobicity (i.e., amphipathic
  sequence) between residues provides cooperativity: folding one helix
  helps the formation of the others.
\item The barrier associated with a two-state transition is
  interpreted in the hydrophobic collapse model as the result of the
  cost of breaking hydrophobic contacts from a maximally compact state
  into the folded ensemble \cite{dill_stigter}.  A further discussion
  on the order of appearance of secondary vs.\ tertiary structure
  formation can be found in the Supporting Material.
\end{enumerate}
Experimental studies of $\alpha$3D showed a fast folding rate of
$(1-5)\,\mu$s and single-exponential kinetics \cite{a3d_kinetics},
compatible with a two-state cooperative transition. As presented here,
this highlights the interplay between secondary structure formation
(see Fig.~\ref{fig:2d_hel_a3d}) and the \emph{loss} of non-native
tertiary contacts (see Fig.~\ref{fig:1d_tert_a3d})---both occurring
exactly within the coexistence region---as a possible mechanism for
folding cooperativity \cite{bereau_jacs}.

Compaction of the unfolded state upon temperature increase has been
observed experimentally by Nettels \emph{et al.}\ using
single-molecule FRET \cite{schuler}. While in our simulations the
decrease in the radius of gyration can be explained by a combination
of the hydrophobic effect and the loss of helical structure, Nettels
\emph{et al.}\ showed similar behavior also for an intrinsically
disordered hydrophilic protein, where other mechanisms likely play a
role.

The present work avoided any reference to free energy barriers so far.
While the nature of the finite-size transition can unambiguously be
characterized from the presence of a convex intruder in the entropy
$S(E)$ \cite{gross_book}, which implies a non-zero latent heat $\Delta
Q$, the mere existence of a free energy barrier is not a strong
criterion because, first, the definition of a free energy barrier is
not unique in a finite-size system \cite{Janke1998, Borgs1992} and,
second, the height of the barrier depends on the reaction coordinate
used. Chan \cite{Chan_coop} therefore argued that the calorimetric
criterion, which relates the van't Hoff and calorimetric energies, is
often more restrictive on protein models than the existence of such a
free energy barrier. Still, the density of states calculations
performed here correlate well with calorimetric ratios for
(AAQAA)$_n$, $n=\{3,\,7,\,10,\,15\}$ and $\alpha$3D: $\delta = 0.78,
\,0.76, \,0.51, \,0.52$ and 0.78, respectively. These were determined
by analyzing the canonical specific heat curve $C_{\rm V}(T)$ as in
Kaya and Chan \cite{Chan_kappa2} ($\kappa_2$ without baseline
subtraction).  The value $\delta=0.78$ for $\alpha$3D also agrees with
an earlier theoretical calculation of the similar bundle $\alpha$3C
from Ghosh and Dill \cite{ghosh_dill}, who found $\delta=0.72$.

\section{Conclusion}

Replica-exchange MD simulations of an intermediate resolution CG
implicit-solvent peptide model allowed us to accurately determine the
thermodynamics of folding for several helical peptides, without
biasing the force field toward a particular native structure. We
argued that a microcanonical analysis is extremely valuable when
characterizing the energetics and structure of peptides, for two
reasons.  First, an accurate density of states calculation allowed the
unambiguous characterization of the nature of the folding transition;
and second, different order parameters, analyzed as a function of $E$,
have exhibited highly non-monotonic behavior inside the
(first-order-like) transition region. A corresponding canonical
analysis (i.e., as a function of $T$) would not allow us to observe in
such detail many of the abovementioned features around transition
regions.

The results showed that simply elongating the (AAQAA)$_n$ sequence
induced a change in the nature of the transition---from two-state
($n=3,\,7$) to downhill ($n=10,\,15$). This correlated with the number
of helices sampled around the transition region/point which is
indicative of the average number of helix nucleation sites, thus
characterizing the number of distinct melting domains and the
structural diversity of intermediates.  Remarkably, the loss of a
first-order signature still goes along with a potentially misleading
steepening of the helicity as a function of temperature for longer
chains (see Supporting Material).  The bundle $\alpha$3D was found to
be two-state cooperative, in agreement with theoretical models
\cite{ghosh_dill, bahato09}. The analysis of tertiary structure
formation highlighted the influence of the amino acid sequence on the
folding mechanism, using the hydrophobic collapse model as a starting
point.

While previous studies have brought forward the coupling between
secondary and tertiary structure formation for two-state cooperativity
(e.g., \cite{kaya_chan_prl, ghosh_dill, bahato09, bereau_jacs}), we
illustrated here several links between the nature of the transition
and secondary/tertiary structure signatures of folding for realistic
representations of peptide chains. Reaching a thorough understanding
of structure formation in two-state cooperative proteins will provide
insight into the stability of their folded conformation.
Cooperativity improves stability of the folded state by suppressing
the population of intermediates.  Mutations that lower cooperativity
not only decrease stability, they have shown to promote misfolding in
certain cases \cite{booth97}.  The resolution of the CG model provides
a useful compromise between computational efficiency and resolution in
order to access features that were so far only observed in less
realistic lattice models.

\section*{Acknowledgment}

We acknowledge stimulating discussions with K.~Binder, W.~Paul,
B.~Schuler, R.~H.~Swendsen, M.~Taylor, and T.~R.~Weikl.

This work was partially supported by grant P01AG032131 from the
National Institutes of Health. M.B.~thanks the Forschungszentrum
J\"ulich for supercomputer time grants jiff39 and jiff43.
T.B.~acknowledges support from an Astrid and Bruce McWilliams
Fellowship.


\bibliographystyle{unsrt}

\newpage

\begin{table}[htbp]
  \begin{center}
    \begin{tabular}{c|c}
      Peptide & Sequence \\
      \hline
      \hline
      helix $n=3$ & (\texttt{AAQAA})$_3$ \\
      helix $n=7$ &  (\texttt{AAQAA})$_{7}$ \\
      helix $n=10$ &  (\texttt{AAQAA})$_{10}$ \\
      helix $n=15$ &  (\texttt{AAQAA})$_{15}$ \\
      bundle $\alpha$3D & \texttt{M\underline{GSWA EFKQR LAAIK TRL}QA LGGSE}\ldots  \\
      & \texttt{A\underline{ELAA FEKEI AAFES ELQA}Y KGKGN}\ldots \\ 
      & \texttt{PEV\underline{EA LRKEA AAIRD ELQAY R}HN}\hspace{7.5mm}
    \end{tabular}
    \caption{Amino acid sequences of the peptides studied in this work.  The three
      helical regions of the native state (from NMR structure, PDB 2A3D) of the
      helix bundle $\alpha$3D \cite{2a3d} are underlined (as predicted by {\sc
        stride} \cite{stride}).}
    \label{tab:aa_seq}
  \end{center}
\end{table}

\newpage

\section*{List of Figure captions}

\begin{figure*}[htbp]
  \includegraphics[width=\linewidth]{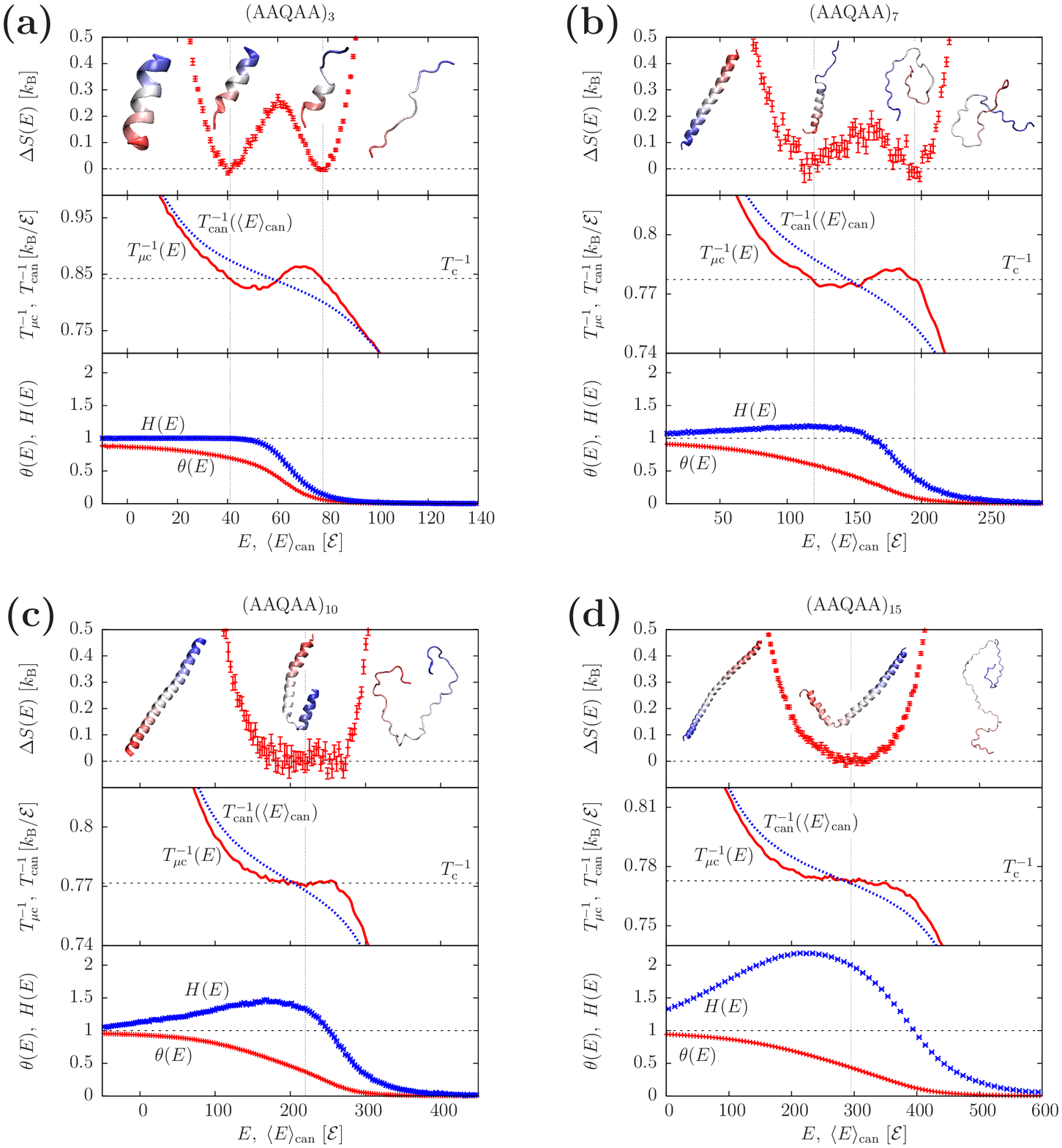}
  \caption{}
  \label{fig:entropy_n}
\end{figure*}

\begin{figure}[htbp]
  \begin{center}
    \includegraphics[height=.9\textheight]{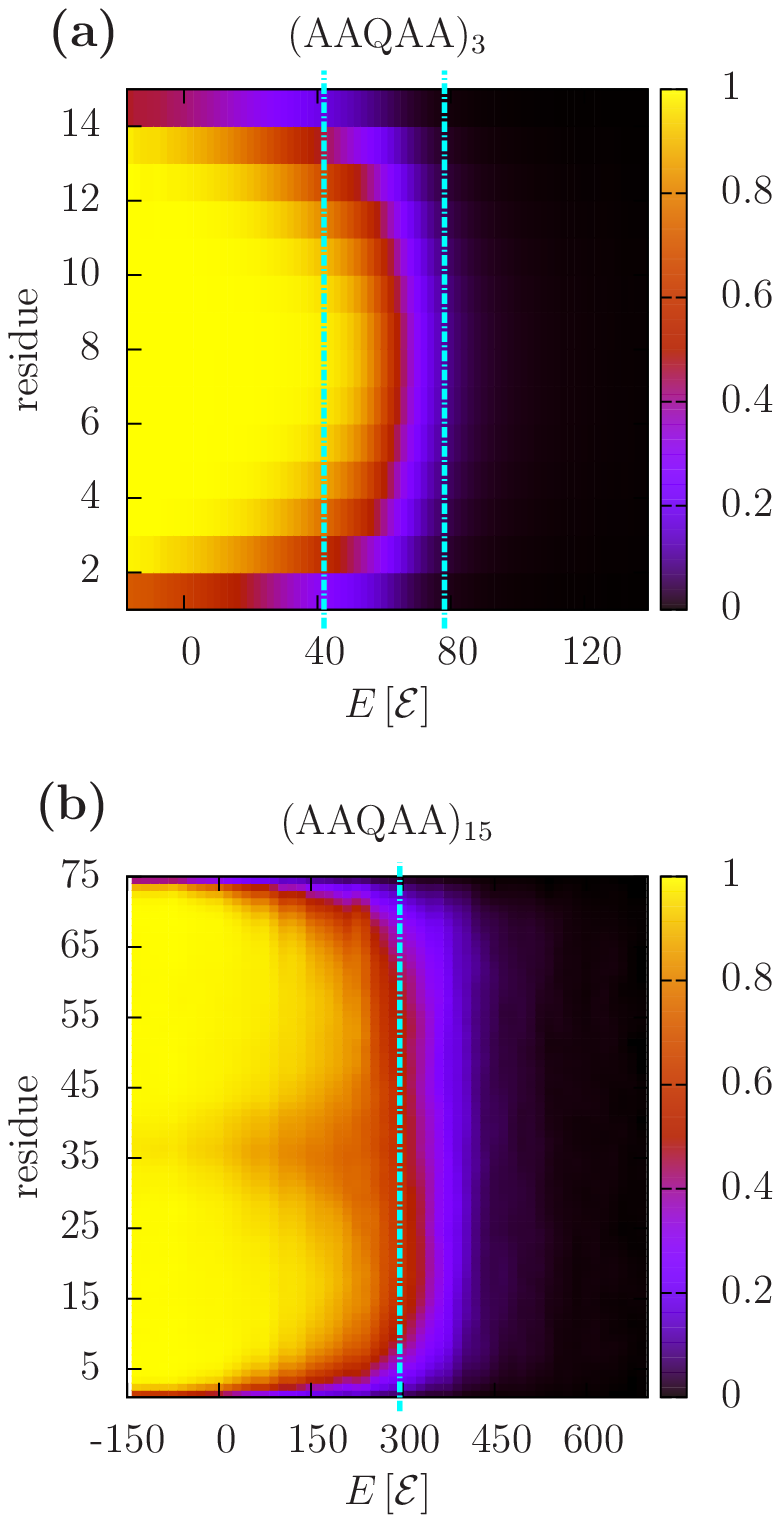}
    \caption{}\label{fig:2d_hel_n}
  \end{center}
\end{figure}

\begin{figure}[htbp]
  \begin{center}
    \includegraphics[width=.7\linewidth]{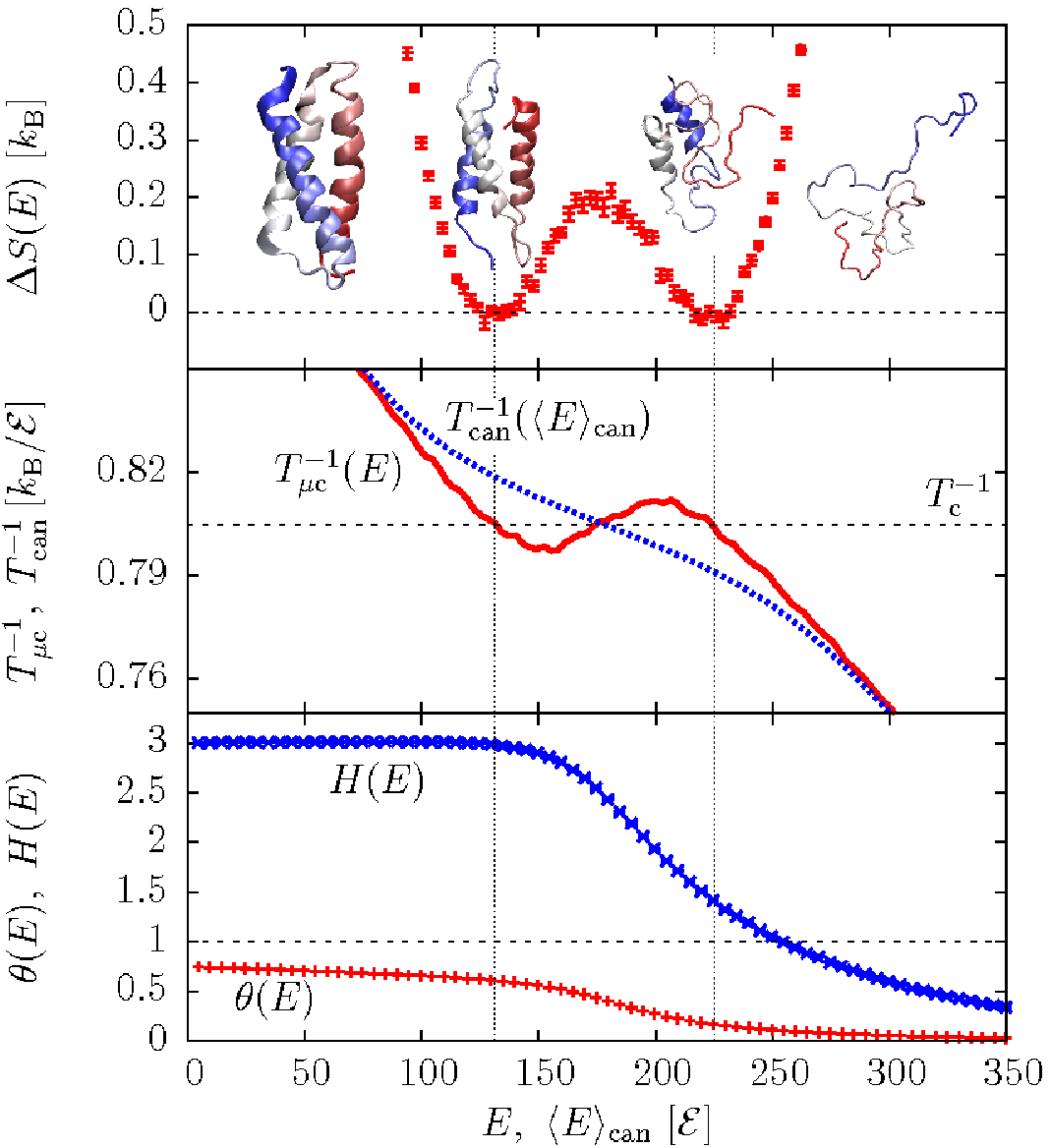}
    \caption{}\label{fig:entropy_a3d}
  \end{center}
\end{figure}

\begin{figure}[htbp]
  \begin{center}
    \includegraphics[width=0.6\linewidth]{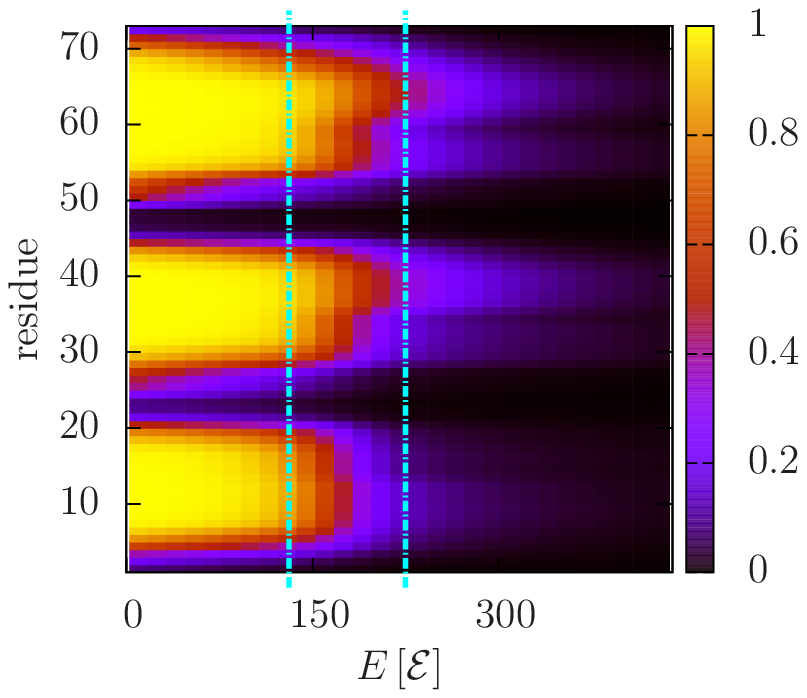}
    \caption{}\label{fig:2d_hel_a3d}
  \end{center}
\end{figure}

\begin{figure}[htbp]
  \begin{center}
    \includegraphics[height=.9\textheight]{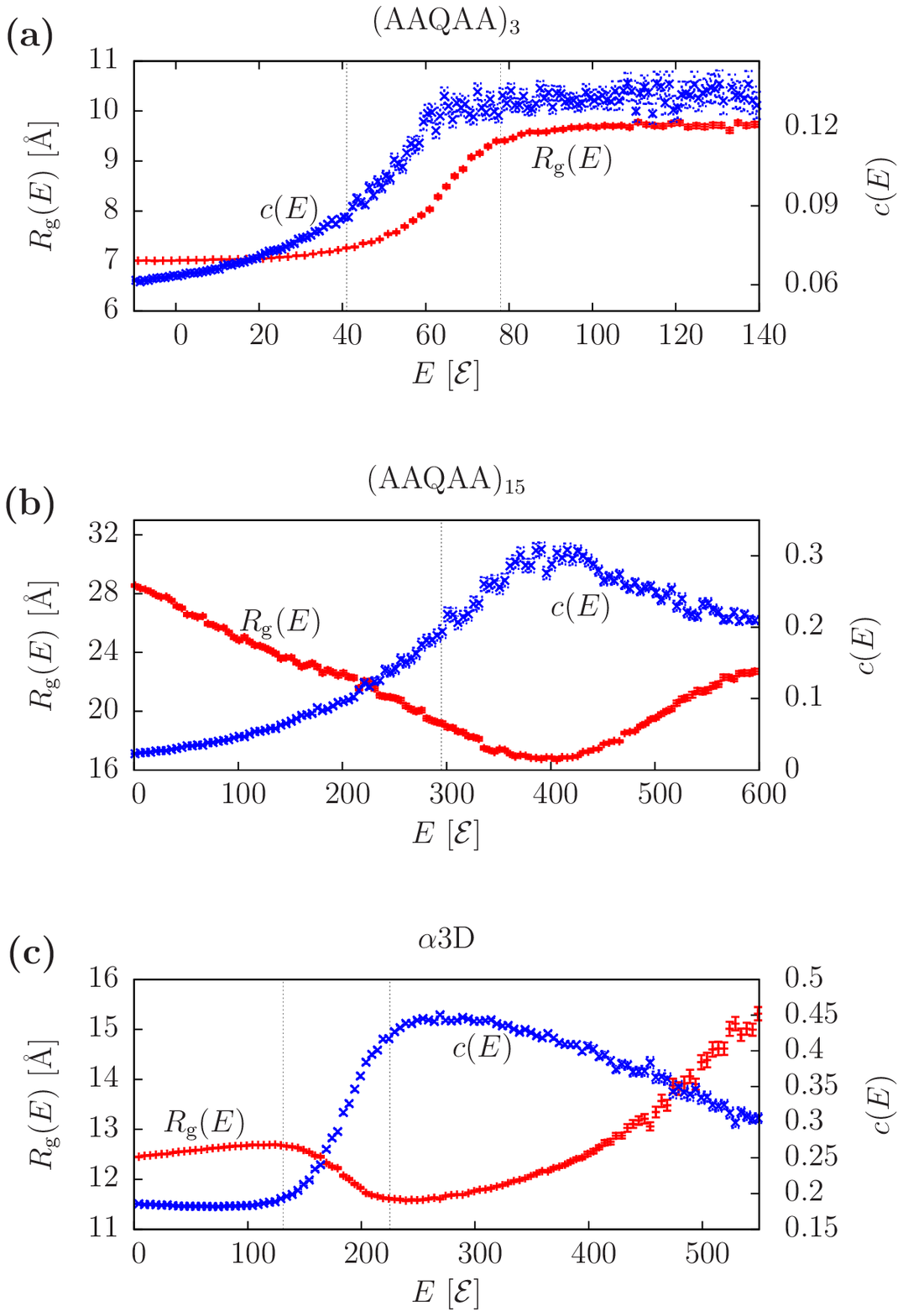}
    \caption{}\label{fig:gyr}
  \end{center}
\end{figure}

\begin{figure}[htbp]
  \begin{center}
    \includegraphics[width=0.8\linewidth]{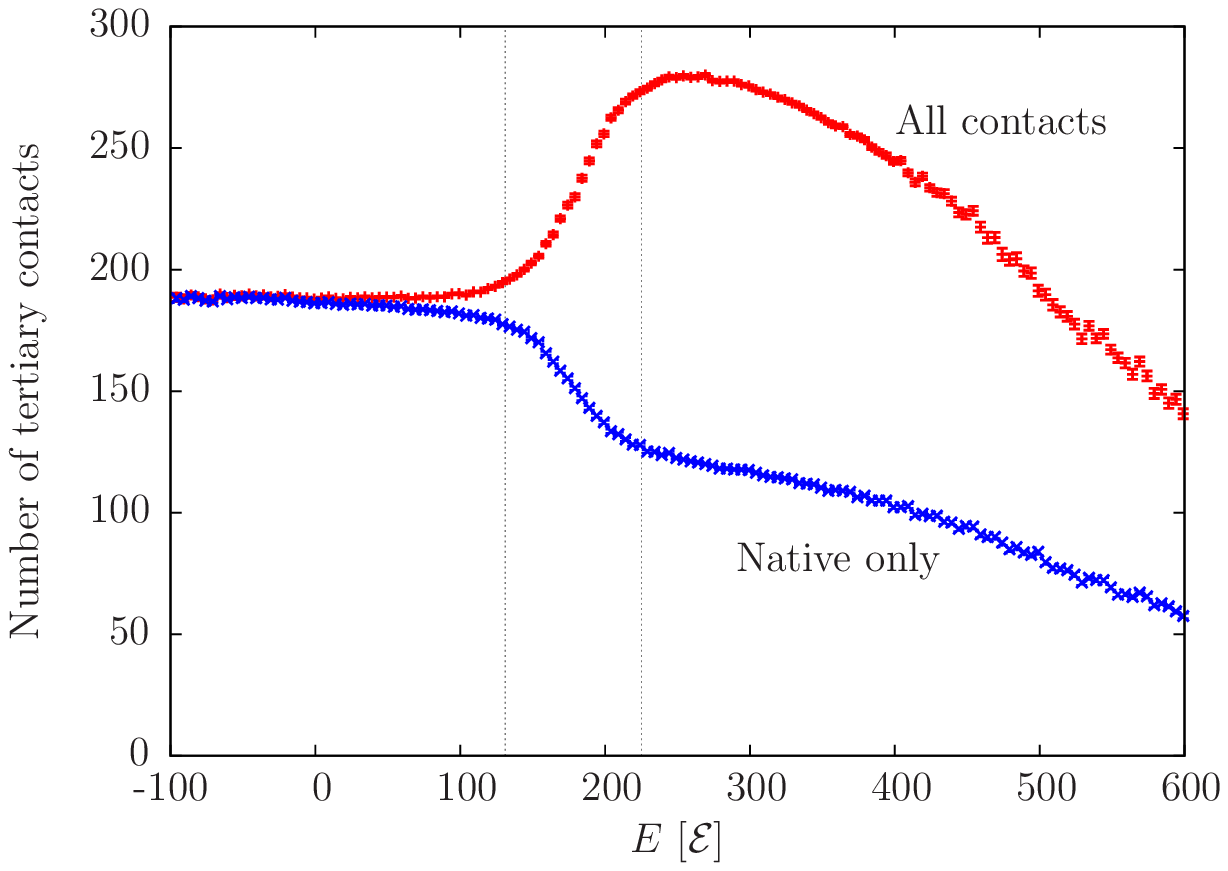}
    \caption{}\label{fig:1d_tert_a3d}
  \end{center}
\end{figure}

\begin{figure}[htbp]
  \begin{center}
    \includegraphics[height=0.9\textheight]{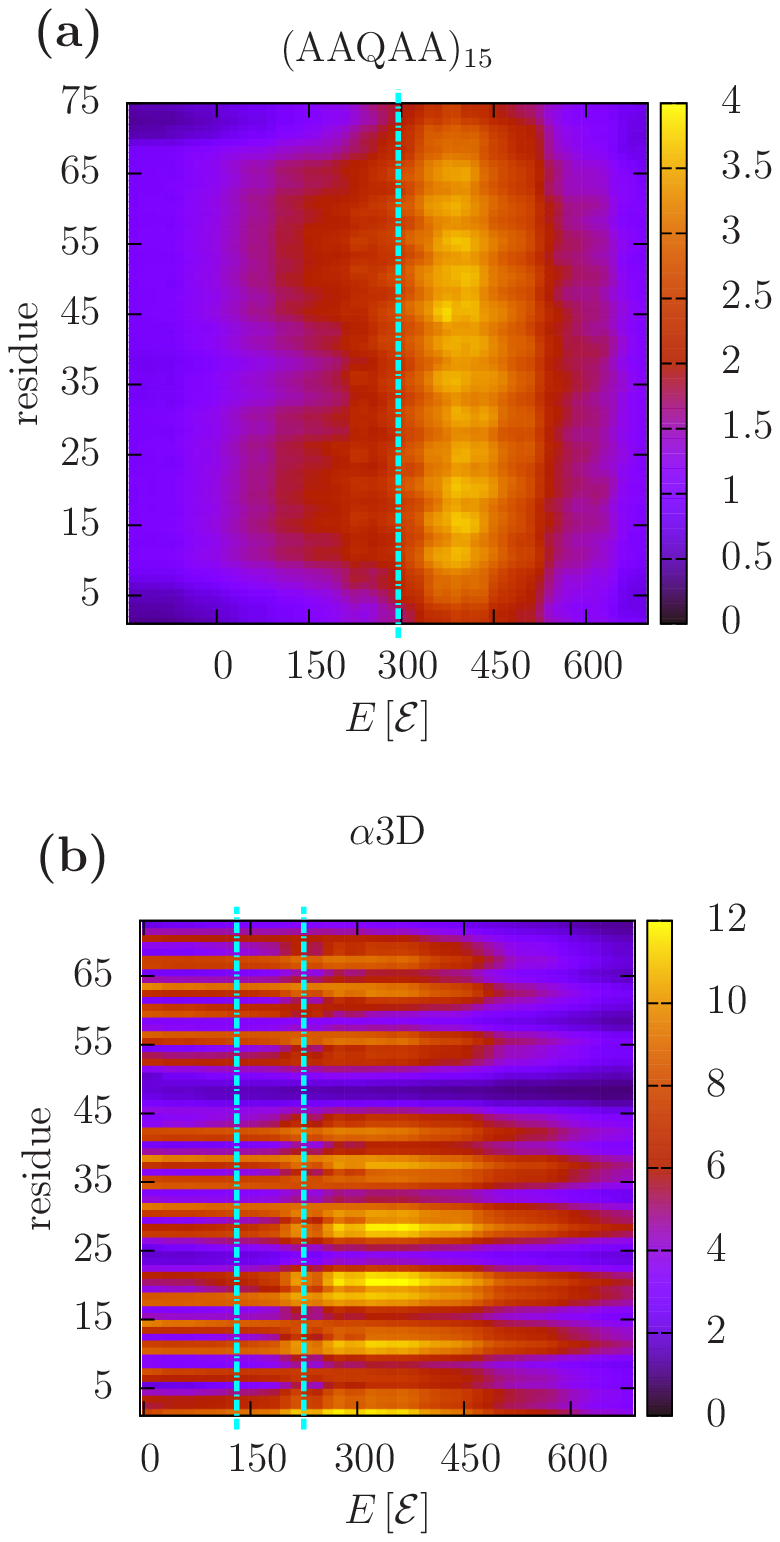}
    \caption{}\label{fig:2d_tert}
  \end{center}
\end{figure}

\begin{description}
\item [Figure 1] Various observables as a function of energy for
  (AAQAA)$_n$: (a) $n=3$, (b) $n=7$, (c) $n=10$, (d) $n=15$. From top
  to bottom for each inset: $\Delta S(E)$, error bars reflect the
  variance of the data points ($1\,\sigma$ interval); inverse
  temperatures from a canonical ($T_{\rm can}^{-1}(\langle E_{\rm
    can}\rangle$, blue) and a microcanonical ($T_{\mu {\rm c}}^{-1} =
  \partial S/\partial E$, red) analysis, where $\langle E_{\rm
    can}\rangle$ is the canonical average energy; helicity $\theta(E)$
  (red) and number of helices $H(E)$ (blue), both with the error of
  the mean. Vertical lines mark either the transition region
  ($n=3,\,7$) or the transition point ($n=10,\,15$).  Representative
  conformations at different energies, visualized using VMD
  \cite{vmd}, are shown.
\item [Figure 2] Fraction of secondary structure as a function of energy and
  residue for (a) (AAQAA)$_3$ and (b) (AAQAA)$_{15}$. Vertical lines mark
  the transition region (a) and point (b), respectively.
\item [Figure 3] Various observables as a function of energy for
  $\alpha$3D. Plots and definitions agree with the conventions in
  Fig.~\ref{fig:entropy_n}.
\item [Figure 4] Fraction of secondary structure as a function of energy and
  residue for $\alpha$3D. Vertical lines mark the transition region.
\item [Figure 5] Radius of gyration $R_{\rm g}(E)$ (red) and normalized
  acylindricity parameter $c(E)$ (blue), both with the error of the mean, for
  (a) (AAQAA)$_3$, (b) (AAQAA)$_{15}$, and (c) $\alpha$3D. Vertical lines mark
  either the transition region ($n=3$, $\alpha$3D) or the transition point
  ($n=15$).
\item [Figure 6] Number of tertiary contacts for $\alpha$3D as a function of
  energy. The ``All contacts'' curve (red) averages over all non-local pairs
  whereas the ``Native only'' curve (blue) only counts native pairs (see text
  for details). Vertical lines mark the transition region.
\item [Figure 7] Number of tertiary contacts as a function of energy
  and residue for (a) (AAQAA)$_{15}$ and (b) $\alpha$3D. Observe that
  the dynamic range of (b) is four times as wide as that for (a).
  Vertical lines mark the transition point (a) and region (b).

\end{description}

\newpage

\end{document}